\documentclass[11pt,a4paper]{article}
\usepackage{jcappub}
\usepackage{graphicx}

\title{Subdominant Dark Matter sterile neutrino resonant 
production in the light of {\sc Planck}}

\author{L. A. Popa, D. Tonoiu}

\affiliation{Institute of Space Science,\\
Bucharest-Magurele, Ro-077125 Romania}

\emailAdd{lpopa@spacescience.ro, tonoiud@spacescience.ro}

\abstract{ 
Few independent detections of a weak X-ray line at an energy of $\sim$3.5 keV 
seen toward a number of astrophysical sites have been reported.
If this signal will be confirmed to be the signature of decaying DM sterile neutrino 
with a mass of $\sim 7.1$ keV, then the cosmological observables should be consistent with its properties.\\
In this paper we make a coupled treatment of the weak decoupling, 
primordial nucleosynthesis and photon decoupling epochs in the sterile neutrino resonant 
production scenario, including the extra radiation energy density via $N_{eff}$.
We compute the radiation and matter perturbations including
the full resonance sweep solution for $\nu_{\alpha}/{\bar \nu}_{\alpha} \rightarrow  \nu_{s}$ 
flavor conversion in the expanding Universe.\\
We show that the cosmological measurements are in agreement with subdominant Dark Matter 
sterile neutrino resonant production with following parameters (errors at 95\% CL):  mass
$m_{\nu_s}=$6.08 $\pm$ 3.22 keV, mixing angle $\sin^2 2 \theta <$5.61 $\times$10$^{-10}$, lepton number per flavor
$L_4 = 1.23 \pm 0.04$ ($L_{4} \equiv 10^4 L_{\nu_a}$) and sterile neutrino mass fraction 
$f_{\nu_s}< 0.078$.\\
Our results are in good agreement
with the sterile neutrino resonant production parameters
inferred  in Ref. \cite{Abazajian14} from the linear large scale structure
constraints to produce full Dark Matter density.}
\keywords{cosmology: cosmic microwave background, cosmological parameters, early universe, neutrino}

\begin{document}
\maketitle

\begin{flushright}
\end{flushright}

\section{Introduction}
\label{S1}

With about 25\% of the total energy density of the Universe 
represented by Dark Matter (DM) and six basic cosmological parameters, 
the $\Lambda$CDM cosmological standard model is remarkably successful at reproducing the large-scale structure (LSS) of the Universe \cite{Planck1,Planck2}. Despite its successes on large-scales, this model produces too much power at small scales. In general, the observed structures are 
less abundant (``missing satellite problem") \cite{Klypin99,Krav10,Zwaan10,Papa11},
have cored density profiles (``core/cusp problem") \cite{Salucci00,Gentile04,Kuzio11} 
and  are too dense (``too-big-to-fail problem") \cite{Boylan11,Boylan12}
than those predicted by the $\Lambda$CDM model.
Numerous attempts involving collisionless DM 
\cite{Krav10,Gov10,Walker11,Ferrero12,Garrison14} fail to resolve these problems, 
suggesting that DM may exhibit gravitational properties as well.\\ 
The nature of DM remains however unknown and the explanation of its existence requires the revision of Standard Model (SM) of particle physics. In this regard, many extensions of SM have been considered  \cite{Jung96,Bertone05,Feng10}.
 
Among the highly motivated candidates for DM is the right-handed sterile neutrino \cite{Aba01,Aba02,Boyarsky09,Kusenko09}.  
The minimal extension of SM by the addition of three right-handed sterile neutrinos (the $\nu$MSM), each one corresponding to 
a SM active neutrino flavor, can simultaneously explain the active neutrino oscillations and the DM properties \cite{Asaka1,Asaka2}. Within this theory, the DM sterile neutrino  masses are in the keV range and the mass of the lightest active neutrino is smaller than $10^{-5}$ eV. This excludes the degenerate mass spectra of the active neutrinos and fixes the absolute mass scale of the other two active neutrinos \cite{Boyarsky06}. Moreover, the $\nu$MSM  can also explain  the matter-antimatter asymmetry of the Universe \cite{Asaka2}. 

In the original Dodelson-Widrow (DW) model \cite{Dod94}, 
sterile neutrinos are produced  in the early Universe 
through non-resonant  oscillations (NRP) with active neutrinos in presence of negligible leptonic asymmetry.
As the DW sterile neutrino is a Warm Dark Matter (WDM) particle with non-negligible velocity dispersion, this suppress the matter power spectrum bellow the free-streaming scale, that affects the DM distribution at small-scales.\\ 
For NRP case, the universal Tremaine-Gunn bound \cite{Gunn} on sterile neutrino mass leads to $m_s^{\rm NRP} > 1.8$ keV. Lower limits on sterile neutrino mass have been placed from various observational probes.
%\footnote{Throughout the paper the sterile neutrino mass is quoted at 95\% CL.}. 
The combination of CMB measurements, LSS and Ly-$\alpha$ forest 
power spectra lead to  $m_s^{\rm NRP} > 1.7$ keV with a further 
improvement to $m^{\rm NRP}_s > 3$ keV  when  high-resolution 
Lyman-$\alpha$ power spectra  are considered \cite{Aba06a}. 
By modeling the SDSS Lyman-$\alpha$ forest flux spectrum, a lower limit of 
$m^{\rm NRP}_s \geq 13$ keV have been found \cite{Viel06,Seljak06}.
The same lower limit has been placed by requiring the number of subhalos  
in N-body simulation to be larger than the number of the observed dwarf spheroidal galaxies (dSphs) of the Milky Way \cite{Rico11}.
Using the phase-space densities derived for dwarf satellite galaxies of the Milky Way a recent work \cite{Horiuchi14}
found a lower limit of DW sterile neutrino mass of  $m^{\rm NRP}_s > 2.5$ keV. \\
Sterile neutrino with a keV mass is known to decay 
radiatively into a photon  and an active neutrino, producing a narrow X-ray 
line with an energy $E_c=\frac{1}{2}m_s$ \cite{Aba01}. 
The width of this decay line increases as the fifth power of sterile neutrino mass and as square of its mixing angle, being potentially detectable in various astrophysical DM sources like X-ray background, galaxy clusters, nearby galaxies and our own Milky Way \cite{Boyarsky06}. Bounds on sterile neutrino radiative decay have been used to place upper limits on sterile neutrino mass.
Strong upper limit constraints in the NRP scenario are obtained from  the observations of Andromeda galaxy (M31) by
XMM-Newton ($m^{\rm NRP}_s < 3.5$ keV) \cite{Watson06,Boyarsky08} and Chandra 
($m^{\rm NRP}_s < 2.2$ keV) telescopes \cite{Watson12}.
When combined, the upper and lower limits 
appears to rule out the sterile neutrino NRP scenario.

The DM sterile neutrino resonant production (RP) via resonant Mikheyev-Smirnov-Wolfenstein (MSW) \cite{MS85,Wolf78} conversion of active neutrinos to sterile neutrinos through the Shi-Fuller mechanism \cite{Shi99}
in the presence of leptonic asymmetry has also 
been proposed \cite{Aba01,Aba02,Laine08}. 
Although the SM of particle physics predicts the leptonic asymmetry of the same order as the baryonic asymmetry, $L \approx B\sim 10^{-10}$, there are scenarios in which much larger leptonic asymmetry can be produced \cite{Smith2006,Cirelli2006,Kirilova2013}. \\
The Tremaine-Gunn bound on sterile neutrino mass for the  RP case leads to 
$m_s^{\rm RP} > 1$ keV \cite{Boyarsky08}.  Also, the combined analysis of WMAP5 and Ly-$\alpha$ data indicates a bound of $m^{\rm RP}_s > 2$ keV in the frame of mixed Cold plus Warm Dark Matter (CWDM) model \cite{Viel09}.\\ 
Sterile neutrino produced via RP mechanism was also forecast to have a radiative decay mode that could be detectable by X-ray telescopes as an emission line in spectra of X-ray clusters and galaxies \cite{Aba01,Tucker01}. \\
Other proposed mechanisms are sterile neutrino production via interactions with the inflaton field \cite{Kusenko06,Tkachev06,Bezrukov14} and by the decay of heavy scalar singlets \cite{Merle14a,Merle14b}.

Recently, few independent detections of a weak X-ray emission line at an energy of $\sim$ 3.5 keV have been found
in a stacked XMM-Newton spectrum of 73 galaxy clusters with redshifts in the range 0.01 - 0.35, in Chandra ACIS-I and ACIS-S spectra of Perseus \cite{Bulbul14a} and in the XMM-Newton spectra of Andromeda galaxy and Perseus galaxy cluster \cite{Boyarsky14a}. Evidences of this emission have been also searched in Milky Way with data from Chandra \cite{Riemer} and XMM-Newton \cite{Boyarsky14b,Jeltema}, in deep Suzaku X-ray spectra of the central regions of Perseus, 
Coma, Virgo and Ophiuchus clusters \cite{Loew,Urban}, 
in the stacked XMM-Newton spectra of dwarf spheroidal galaxies in the vicinity of Milky Way \cite{Malyshev} and in the stacked spectra of a sample of galaxies selected from Chandra and XMM-Newton public archives\cite{Anderson}. 
%Also, a re-analysis of XMM-Newton data of M31 was done in \cite{Jeltema}.
Until now, none of these searches were able to establish precisely if the origin of this line is is a signature of decaying dark matter \cite{Boyarsky14a} or it is
related to any known atomic transition in thermal plasma, with special 
interest in Potassium and Chlorine atomic transitions \cite{Riemer,Jeltema,Boyarsky14b}.  
The evidence (or lack of evidence in some cases) of this emission line in different astrophysical sites placed strong constraints on both hypothesis, raising several controversies on this subject \cite{Boyarsky14c,Bulbul14b,Carlson}.\\
Sterile neutrinos produced through a resonant mechanism 
remains however the simplest model for DM origin of the undefined $\sim$3.5 keV of X-ray line. 
The RP sterile neutrino parameters required to produce this signal  
have been recently inferred from the linear large scale structure
constraints to produce full DM density \cite{Abazajian14}.
These parameters are consistent with the Local Group and high-z galaxy count 
constraints, fulfilling previously determined requirements to successfully 
solve  ``missing satellite problem"and ``too-big-to-fail problem".

The cosmological constraints on subdominant RP sterile neutrino parameters 
have not yet derived. The goal of this paper is to place constraints on
RP sterile neutrino parameters in a $\Lambda$CWDM model containing a mixture of 
WDM in the form of RP sterile neutrino and CDM, by using most of the present cosmological measurements.\\
In particular, the reconstructed gravitational lensing potential power spectrum obtained 
by the {\sc Planck } satellite has impact on cosmological parameter degeneracies 
when DM sterile neutrino  scenario is considered \cite{Popa07}, 
offering several advantages  against other methods \cite{Planck2,Perotto13}. 
As the gravitational lensing effect depends on the DM distribution in the Universe, 
no assumption on light-to-mass bias is required. The projected gravitational lensing potential is sensitive to the matter distribution out to high redshifts, preventing from non-linear corrections required only at very small scales. In addition, unlike the galaxy clustering and the Ly-$\alpha$  forest, the projected gravitational lensing potential probes a larger range of angular scales, most of the signature coming from large scales. 
Moreover, the preference of present cosmological data for a small but significant leptonic asymmetry 
have been recently found \cite{Popa14}. \\
This paper is organized as follows. In Sec. 2 we briefly review the DM sterile neutrino RP 
calculations. In Sec. 3 we describe the model, methods used in our analysis and the datasets we consider. We present our results in Sec. 4 where we examine the consistency and cosmological implications of DM sterile neutrino RP scenario. In Sec. 5 we draw our conclusions.

\section{Sterile neutrino resonant production calculations} 
\label{S2}

The DM sterile neutrinos  with rest mass $m_s \geq$ 1 keV  are produced through neutrino oscillations at temperatures close to the QCD phase transition \cite{Aba01,Aba02}. The oscillation process take place because
the neutrino mass eigenstate components propagate differently  
as they have different energies, momenta and masses. Eigenstates of neutrino interaction include the active neutrinos/antineutrinos, 
$\nu_{\alpha}$/${\bar \nu}_{\alpha}$
($\alpha$=$e$, $\mu$, $\tau$), which are created and destroyed in
the SM of particle physics by weak interactions as well as the  sterile neutrinos/antineutrinos, $\nu_s$/${\bar \nu}_s$, 
which do not participate in weak interactions. \\
For the purpose of this work we consider three active neutrino
flavors $\nu_{\alpha}/{\bar \nu}_{\alpha}$  and one sterile neutrino $\nu_{s}/{\bar{\nu}_s}$ with the production channel 
$\nu_{\alpha}$/${\bar \nu}_{\alpha}$ $\rightarrow$ $\nu_s$/${\bar \nu}_s$. 
 In the early Universe, this conversion 
can occur through the medium-enhanced MSW resonant mechanism in presence of 
an initial neutrino lepton number residing in the active neutrinos \cite{Aba01,Aba02,Shi99}:
\begin{eqnarray} 
\label{init_lepton}
L_{\nu_{\alpha}} \equiv \frac{( n_{\nu_{\alpha}} - n_{{\bar \nu_{\alpha}})}} {n_{\gamma}}\,,
\end{eqnarray}
where $n_{\nu_{\alpha}}$/$n_{{\bar \nu_{\alpha}}}$ and ${n_{\gamma}}$ are the 
active neutrino/antineutrino and  photon number densities respectively, 
$n_{\gamma}=2\zeta(3)T^3_{\gamma}/\pi^2$ and $\zeta(3)$ is Riemann zeta function of 3. \\
Above the neutrino weak decoupling ($T_{dec} \sim 1$ MeV), the weak interaction rates of neutrino scattering are much faster than the Hubble expansion rate and  
the neutrino energy distribution can reshuffle into a thermal Fermi-Dirac spectrum. 
After $T_{dec}$, when active neutrinos decouple from the thermal plasma, 
the conversion process can be resonantly enhanced \cite{Aba05},
affecting the primordial nucleosynthesis ($T \sim 0.1$ MeV) and 
the photon decoupling ($T \sim 0.2$ eV) epochs \cite{Grohs14,Grohs15}.\\
In this section we follow the detailed resonant production calculations first 
performed in Refs. \cite{Aba01,Aba05} for 
adiabatic and continuously sweeping active-sterile 
resonance, extended  then to the non-adiabatically sweeping case 
in Refs. \cite{Kishimoto06,CSmith} for the computation of the  primordial nucleosynthesis abundance yields (the full solution) and 
examine how the sterile neutrino resonant production affects   
the photon decoupling.\\

\subsection{Lepton number depletion and evolution of resonance energies}
\label{S2.1}

The MSW condition for the resonant scaled neutrino momentum $\epsilon_{res}=p/T|_{res}$ 
is given by:
\begin{equation}
\label{eps_res}
\epsilon_{res} \approx \left( \frac{\delta m^2 \cos 2\theta}
{4 \sqrt{2} \zeta(3)\pi^{-2}G_F {{\cal L_{\nu_{\alpha}}}}} \right) T^{-4}\,, 
\end{equation}
where $\delta m^2=m_2^2-m_1^2\approx m_2^2$ is the difference of the squares of  sterile neutrino and active neutrino eigenvalues, $T$ is the photon/plasma temperature,
$G_F$ is the Fermi constant, $\theta$ is the vacuum mixing angle and ${\cal L_{\nu_{\alpha}}}$ is the potential lepton number corresponding to an active neutrino flavor $\alpha$:
\begin{eqnarray}
{\cal L_{\nu_{\alpha}}} \equiv 2 L_{\nu_{\alpha}} +\sum_{\beta \ne \alpha} L_{\nu_{\beta}} \,,
\hspace{0.3cm}\beta=(e, \mu, \tau)\,. 
\end{eqnarray}
%The individual lepton number $L_{\nu_{\alpha}}$ is given in terms of neutrino, antineutrino %and photon number densities as $L_{\nu_{\alpha}} \equiv ( n_{\nu_{\alpha}} - n_{{\bar %\nu_{\alpha}})}/ n_{\gamma}$. 
The leptonic asymmetry  is most conveniently measured by the neutrino degeneracy parameter \cite{Dolgov,Serpico,Simha} defined as
$\xi_{\nu_{\alpha}}=\mu_{\nu_{\alpha}} /T_{\nu}$, where $\mu_{\nu_{\alpha}}$ is the neutrino chemical potential and $T_{\nu}$ is the present temperature of the neutrino background [$T_{\nu}/T_{\gamma}=(4/11)^{1/3}$]. The relation between the lepton number of  neutrino flavor $\alpha$ and the corresponding chemical potential is given by:
\begin{eqnarray}
L_{\nu_{\alpha}} = \left( \frac{1}{12 \zeta(3)} \right)
\left( \frac{T_{\nu}}{T_{\gamma}}\right)^3
[\pi^2 \xi_{\nu_{\alpha}} + \xi^{3}_{\nu_{\alpha}} ] \,.
\end{eqnarray}
The evolution of the neutrino flavor asymmetries with the temperature 
is driven by the active neutrino oscillation mixing parameters \cite{Wong,Abz}.
These parameters are accurately measured with the exception of one mixing angle, $\theta_{13}$, that only recently started to be constrained.
Actually, the value of $\theta_{13}$ plays a decisive role in determining the temperature at which the equilibrium among active neutrino flavor asymmetries is reached
\cite{Mangano12,Casto12}. For values of $\theta_{13}$ closed to the present experimental upper bound, it is shown that the active neutrino oscillations are effective at a temperature $T \sim$ 8 MeV \cite{Mangano11} and therefore the active neutrinos reach the chemical equilibrium before BBN. For this reason we consider in the present work the same  degeneracy parameter for all neutrino flavors: $\xi_{\nu_e}=\xi_{\nu_{\mu}}=\xi_{\nu_{\tau}}$.\\
As the universe expands, the temperature falls
causing the resonance to sweep from low to higher values of the resonant scaled 
neutrino momentum $\epsilon_{res}$. This process converts 
$\nu_{\alpha}$/${\bar \nu}_{\alpha}$ to $\nu_s$, reducing the potential lepton number ${\cal L_{\nu_{\alpha}}}$. 
As discussed in Refs.\cite{Shi99,Aba01}, the evolution of 
${\cal L_{\nu_{\alpha}}}$ is dictated by the resonance sweep rate and the dimensionless adiabaticity parameter $\gamma$ given by:
\begin{eqnarray}
\label{lz}
\gamma  \approx  \frac{\sqrt{5}\zeta(3)^{3/4}}{2^{1/8}\pi^3}
   \frac{(\delta m^2)^{1/4}m_{pl}G^{3/4}_F}{g^{1/2}}
\frac{\sin^2 2 \theta}{\cos^{7/4} 2 \theta} {\cal L_{\nu_{\alpha}}}^{3/4} \epsilon^{-1/4} 
 \times  \left| 1+\frac{{\dot g}/g}{3H} - \frac{\dot{\cal L_{\nu_{\alpha}}}
/{\cal L}_{\nu_{\alpha}}}{3H} \right|^{-1} \,,
\end{eqnarray}
where $m_{pl}$ is the Planck mass, $H$ is the Hubble expansion rate and $g$ is the total statistical weight for relativistic species in the early Universe. Large values of $\gamma$ ($\gamma > 1$) result when the oscillation length is smaller than the resonance width, leading to efficient flavor transformation. If the time rate change of ${\cal L}_{\nu_\alpha}$ is larger than
the Hubble expansion rate, the evolution is non-adiabatic with $\gamma < 1$. 
The likelihood probability of a neutrino at resonance to make the jump between the mass eigenstates is given by Landau-Zerner probability, $P_{LZ}=exp(-\pi \gamma /2)$ \cite{LZ}. 
The evolution of the potential lepton number when the resonant active neutrino/antineutrino  momentum sweeps from 0 to $\epsilon_{res}$ is then given by:
\begin{eqnarray} 
\label{lpot}
{\cal L}_{\nu_{\alpha}}(\epsilon_{res})={\cal L}^{init}-\frac{1}{2 \zeta(3)}\left( \frac{T_{\nu}}{T_{\gamma}}\right)^3 
\int^{\epsilon_{res}}_{0}  (1-e^{-\pi \gamma/2}) 
(f_{\nu_{\alpha}} + f_{{\bar \nu}_{\alpha}}) {\rm d}x \,,
\end{eqnarray}
where $f_{\nu_{\alpha}}$ and $f_{{\bar \nu}_{\alpha}}$ are the initial neutrino/antineutrino  
Fermi-Dirac distribution functions:
\begin{eqnarray}
\label{FD}
f_{\nu_{\alpha}}(\epsilon)  =  \frac{1}{F_2 (\xi_{\nu_{\alpha}})}
\frac{\epsilon^2}{e^{\epsilon-\xi_{\nu_{\alpha}}}+1}\,,
\hspace{0.5cm} 
f_{{\bar \nu}_{\alpha}}(\epsilon) \frac{1}{F_2 (-\xi_{\nu_{\alpha}})}
\frac{\epsilon^2}{e^{\epsilon+\xi_{\nu_{\alpha}}}+1}\,,  
\end{eqnarray} 
and  $F_2 (\pm \xi_{\nu_{\alpha}})$ are the relativistic Fermi integrals of second order: 
\begin{equation}
F_2 (\pm \xi_{\nu_{\alpha}})  \equiv  
\int^{\infty}_0\frac{ x^2 }{e^{x \mp \xi_{\nu_{\alpha}}}+1} {\rm d}x \,. 
\end{equation}

\subsection{Phase-space distributions }
\label{S2.2}

The time evolution of the sterile neutrino phase-space distribution, $f_{\nu_s}$, can be described by the classical Boltzmann equation \cite{Aba01,Aba02}: 
\begin{eqnarray}
\label{bol}
\frac{\partial }{\partial t}f_{\nu_s}(\epsilon)-H(t)\,\epsilon\,\frac{\partial}
{\partial \epsilon}f_{\nu_s}(\epsilon) 
\approx \Gamma(\nu_{\alpha} \rightarrow \nu_s)[f_{\nu_{\alpha}}(\epsilon)-f_{\nu_s}(\epsilon)]\,, 
\end{eqnarray}   
where: $f_{\nu_{\alpha}}$ is the active neutrino phase-space distribution, $H(t)$ is the Hubble expansion rate and $\Gamma(\nu_{\alpha} \rightarrow \nu_s)$
is the sterile neutrino effective production rate:
\begin{equation}
 \Gamma(\nu_{\alpha} \rightarrow \nu_s)\approx 
\frac{1}{2}\Gamma_{\nu_{\alpha}}(p,T) 
\left < P_m(\nu_{\alpha} \rightarrow \nu_{s}) \right > \,,
\end{equation} 
where  $\Gamma_{\nu_{\alpha}}$ is the collision rate:
\begin{displaymath}
\Gamma_{\nu_{\alpha}}(p,T)\approx 
\left\{
\begin{array}{ll}
1.27 G^2_F p T^4, & \alpha =e\\
0.92 G^2_F p T^4, & \alpha =\mu,\tau 
\end{array} \right. 
\end{displaymath}
%The sterile antineutrino phase-space distribution, $f_{{\bar \nu}_s}$, is obtained 
%by performing the transformation $\nu_{\alpha} \rightarrow {\bar \nu}_{\alpha}$
%and $\nu_s \rightarrow {\bar \nu}_s$ in the above equations. \\
The average oscillation probability 
$\left< P_m ( \nu_{\alpha} \rightarrow \nu_{s} ) \right>$ is given by: 
\begin{equation}
\label{sin2_m}
\left< P_m(\nu_{\alpha} \rightarrow \nu_{s} ) \right >
 = \frac{1}{2} \frac{\Delta^2(p) \sin^2 2\theta}
{\Delta^2(p) \sin^2 2\theta+ D^2(p)+
[ \Delta(p)\cos 2\theta+V^{L}-V^{T}] ^2}
\end{equation}
where $\Delta(p)=\delta m^2/2 p$ is the vacuum oscillation factor, $D(p)=\Gamma_{\nu_{\alpha}}(p)/2$ is the quantum damping rate, $V^T$ is the thermal potential and $V^L$ is the asymmetric lepton potential: 
\begin{eqnarray}
V^{L}=2 \sqrt{2}\zeta(3)\pi^{-2}G_FT^3{\cal L}_{\alpha} \,.
\end{eqnarray}
For temperatures characteristic to the post weak decoupling era (T $<$ 3 MeV) 
the contribution of the thermal potential $V^{T}$ in Eq. (\ref{sin2_m}) is very small and can be neglected. \\
Starting from $\nu_{\alpha}$/${\bar \nu}_{\alpha}$ thermal distributions 
at a scale factor $a/a_0 \sim 10^{-13}$ ($a_0\equiv$1 today), corresponding to 
a temperature $T \simeq {\rm 133 MeV} (m_s/1 {\rm keV})^{1/3}$ 
(near sterile neutrino maximum rate production temperature),
we simultaneously evolve  Eqs. (\ref{eps_res}), (\ref{lz}), (\ref{lpot}) and (\ref{bol}) 
%together with the time-temperature relation \cite{Aba01,Popa07} 
to obtain the evolution of ${\cal L}_{\nu_{\alpha}}(\epsilon_{res})$  
in the expanding Universe for the entire range of resonant scaled neutrino momentum values. 
In this calculation we neglect the effects of the QCD quark-gluon phase transition that may 
disturb the DM sterile neutrino production rate by the alteration of the time-temperature relationship in 
the early Universe (due to variation of the statistical weight in different particles) \cite{Aba01,Popa07}.\\ 
Figure~1 presents the dependence of $\epsilon_{res} {\cal L}_{\nu_e}(\epsilon_{res})$  
on sterile neutrino masses and mixing angles for two different initial lepton numbers, 
$L_{4}=$12 and 24 (we define $L_{4}\equiv 10^4 L_{\nu_{\alpha}}$). 
The plots presented show that the maximum on the $\epsilon_{res} {\cal L}_{\nu_e}(\epsilon_{res})$ 
curves occurs at different scaled neutrino momentum values $\epsilon_{max}$. 
For each case, the resonance sweeps adiabatically and continuously 
up to $\epsilon_{max}$. For $\epsilon_{res}> \epsilon_{max}$ the resonance sweeps 
non-adiabatically resulting in an incomplete conversion $\nu_{\alpha} \rightarrow \nu_s$ 
and a partial depletion of ${\cal L}_{\nu_{\alpha}}$. \\ 
Figure~2 presents the dependence of the final phase-space distribution of sterile neutrino $f_{\nu_s}(\epsilon)$  
on the comoving momentum $\epsilon=p/T$ obtained for $\nu_e \rightarrow \nu_{s}$ flavor transformation.
%Pana aici 
The models with fixed $m_s$ and $\sin^2 2\theta$ and increasing $L_4$ values have increasing average of the comoving momenta $\left< p/T \right>$, leading to larger-scale cutoffs in the gravitational potential and matter power spectra.
The same behavior is present for models with fixed $L_4$ and 
$\sin^2 2\theta$ and increasing $m_s$ values. 
The models with fixed $L_4$ and $m_s$ and increasing 
$\sin^2 2\theta$ values have opposite behaviors.
Larger values of $\sin^2 2\theta$ lead to larger sterile neutrino production rates. 
The resonance through momentum distribution occurs in this case at higher temperatures 
and smaller comoving momenta, as can be seen from Eqs.(2.2). \\
For this computation we assume a background cosmology consistent with the 
most recent cosmological measurements \cite{Planck2} with energy 
densities of $\Omega_{m}=0.28$ in matter, $\Omega_{\Lambda}=0.72$  in 
cosmological constant, a Hubble parameter of $H_0$= 69 km s$^{-1}$Mpc$^{-1}$ 
and a total active neutrino mass $m_{\nu}$=0.35 eV,  
as indicated by the minimal extension of the base $\Lambda$CDM model by the addition 
of three degenerated active neutrino flavors (see model M1 below).
\begin{figure}
\label{epsL}
\begin{center}
\includegraphics[height=8cm,width=16cm]{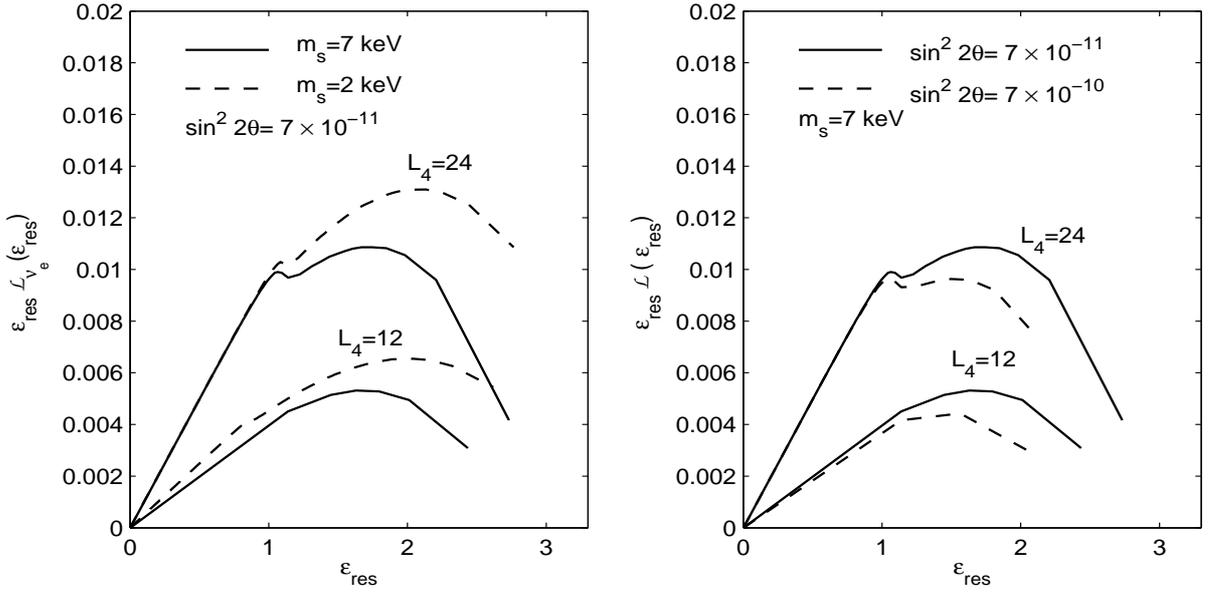}
\end{center}
\caption{Left: The dependence of $\epsilon_{res} {\cal L}_{\nu_e}(\epsilon_{res})$  
on sterile neutrino mass $m_s$=7 keV (continuous line) and $m_s$=2 keV (dashed line) for initial lepton numbers $L_{4}=$ 12, 24  and sterile neutrino mixing angle 
$\sin^2 2\theta=7 \times 10^{-11}$.
Right: The dependence of $\epsilon_{res} {\cal L}_{\nu_e}(\epsilon_{res})$  
on sterile neutrino mixing angle $\sin^2 2\theta $=7 $\times 10^{-11}$ (continuous line) and 
$\sin^2 2\theta=7 \times 10^{-10}$ (dashed line) 
for initial lepton numbers $L_{4}=$ 12,24  and sterile neutrino mass  $m_s$=7 keV.
We define $L_{4}\equiv 10^4 L_{\nu_{\alpha}}$.}
\end{figure}
\begin{figure}
\label{phase}
\begin{center}
\includegraphics[height=8cm,width=16cm]{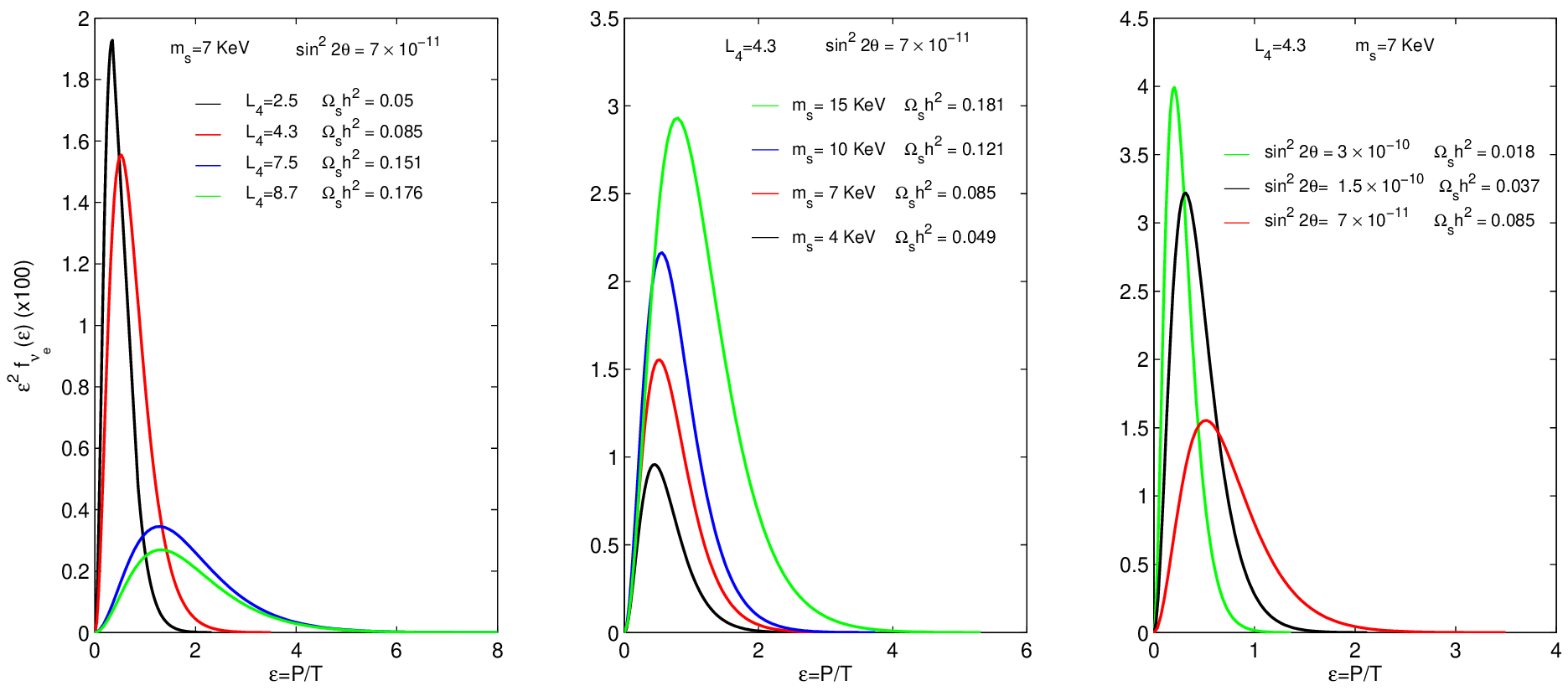}
\end{center}
\caption{The dependence of the final phase-space distribution of sterile neutrino $f_{\nu_s}$  
on the comoving momentum $\epsilon=p/T$ obtained for $\nu_e \rightarrow \nu_{s}$ flavor transformation. From top to bootom 
the plots represent sterile neutrino phase-space distributions of: 
$m_s$=7 keV and $\sin^2 2\theta=7 \times 10^{-11}$ models with $L_4$=2.5,4.3,7.5 and 8.7 (left panel), $L_4$= 4.3 and $\sin^2 2\theta=7 \times 10^{-11}$ models with 
$m_s$=4,7,10 and 15 keV (middle panel) and $L_4$= 4.3 and $m_s$=7 keV models with 
$\sin^2 2\theta$= 3 $\times$10$^{-10}$, 1.5 $\times$10$^{-10}$ and $7 \times 10^{-11}$. For each case 
we also indicate the sterile neutrino energy density parameter, $\Omega_s$h$^2$ at the present time.}
\end{figure}

\subsection{Effects of DM sterile neutrino resonant production on $N_{eff}$}

\begin{figure}
\label{rs}
\begin{center}
\includegraphics[height=9cm,width=17cm]{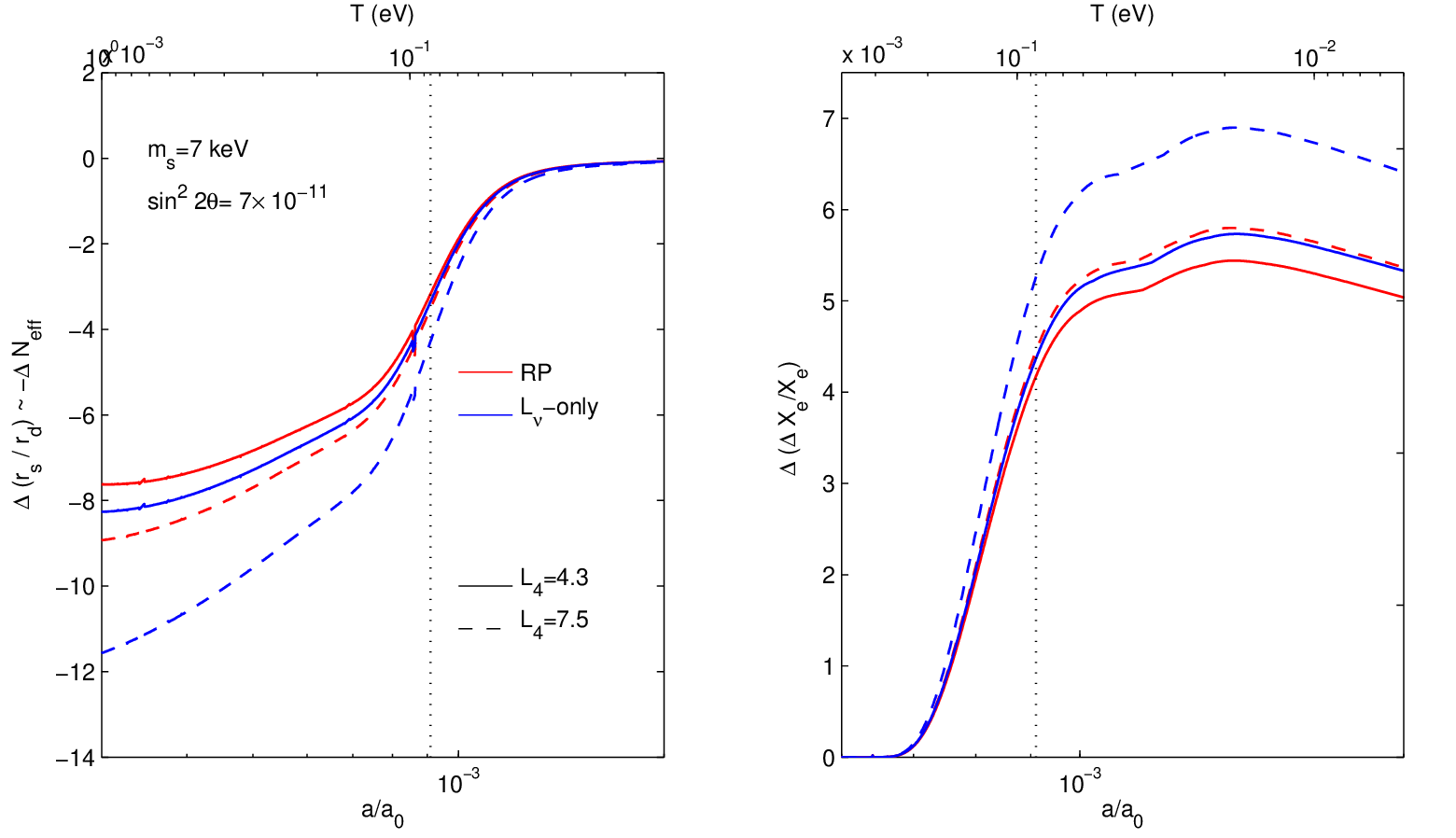}
\end{center}
\caption{ Evolution with the scale factor of the changes in the effective number of relativistic degrees of freedom, $N_{eff}$, 
(left) in and of the free electron fraction, $X_e$,  (right) 
for two lepton number values, $L_4$=4.3 (continuous lines) and $L_{4}$=7.5 (dashed lines), relative to the case with zero lepton number,
for sterile neutrino resonant production models (red lines - RP) and 
and leptonic asymmetric models (blue lines - $L_{\nu}$ only).
The vertical dotted line corresponds to the epoch of photon decoupling 
($a_{dc} \simeq 9 \times 10^{-4}$).}
\end{figure}
We examine the impact of DM sterile neutrino resonant production on the effective number of relativistic degrees of freedom, $N_{eff}$, parameter related to the radiation energy density at the photon decoupling.
We infer $N_{eff}$ from the value of $r_s/r_d$, the ratio  
of the sound horizon $r_s$ to the photon diffusion length $r_d$ at the photon decoupling,
parameter proposed to be used as a proxy for $N_{eff}$
in the case of cosmological models that involve non-equilibrium neutrino distributions \cite{Grohs14,Grohs15}. \\
A simple scaling analysis  \cite{Grohs14} show that a larger Hubble rate results 
in a smaller value of the ratio $r_s/r_d$ and since $\Delta(r_s/r_d)\sim -\Delta N_{eff}$   
($r_s/r_d$ is monotonically decreasing with $N_{eff}$) this imply that a larger Hubble rate leads to a larger value of $N_{eff}$.\\
Employing the same background cosmology as in previous subsection 
and a primordial helium mass fraction $Y_P$=0.24,
we compute the evolution with the scale factor of $r_s$ and $r_d$ 
given by the following integrals:
\begin{flushleft}
\begin{eqnarray}
\label{sound}
r_s(a)= \int^a_0 \frac{ {\rm d}a'}{a'^2 H(a')}
\frac{1}{3(1+R)} \,, 
\hspace{0.3cm} 
r^2_{d}(a)= \pi^2 \int^a_0 \frac{{\rm d}a'}{a'^2 H(a')}
\frac{1}{an_e(a')\sigma_T}\frac{R^2+\frac{16}{15}(1+R)}{6(1+R)^2} 
\end{eqnarray}
\end{flushleft}
where $H(a')$ is the Hubble expansion rate, $n_e(a')$ is the free-electron number density, 
$\sigma_T$ is the Thomson cross section and 
$R(a) \equiv 3\rho_b/(4\rho_{\gamma})$ is the 
ratio of the baryon energy density $\rho_b$, to the photon energy density $\rho_{\gamma}$. 
We also use the recombination code {\sc Recfast} \cite{recfast} to evolve 
the free electron fraction $X_e$ and the number density of free electrons $n_e$ 
with the scale factor, required for the computation of $r_d$.\\
In Figure~3 we plot the evolution with the scale factor of the changes in $N_{eff}$ and $X_e$ 
relative to the model with zero lepton number  
for sterile neutrino resonant production models with $L_4$=4.3 and 7.5 (RP-case)
and for the leptonic asymmetric models with the same values of $L_4$ ($L_{\nu}$-only case).
We see a larger value of $N_{eff}$ at the photon decoupling (that imply a higher Hubble expansion rate) in both cases, relative to the 
case with zero lepton number. The increase of $N_{eff}$ is smaller in the RP case  
relative to the $L_{\nu}$-only case with the same of lepton number.\\
Also, a larger Hubble expansion rate at the photon decoupling implies a larger value of 
the free-electron fraction $X_e$ and an earlier epoch of $X_e$ freeze-out.

\section{Cosmological constraints}
\label{S3}

\subsection{Method and datasets}
\label{S3.1}

\begin{figure}
\label{phase}
\begin{center}
\includegraphics[height=8cm,width=14cm]{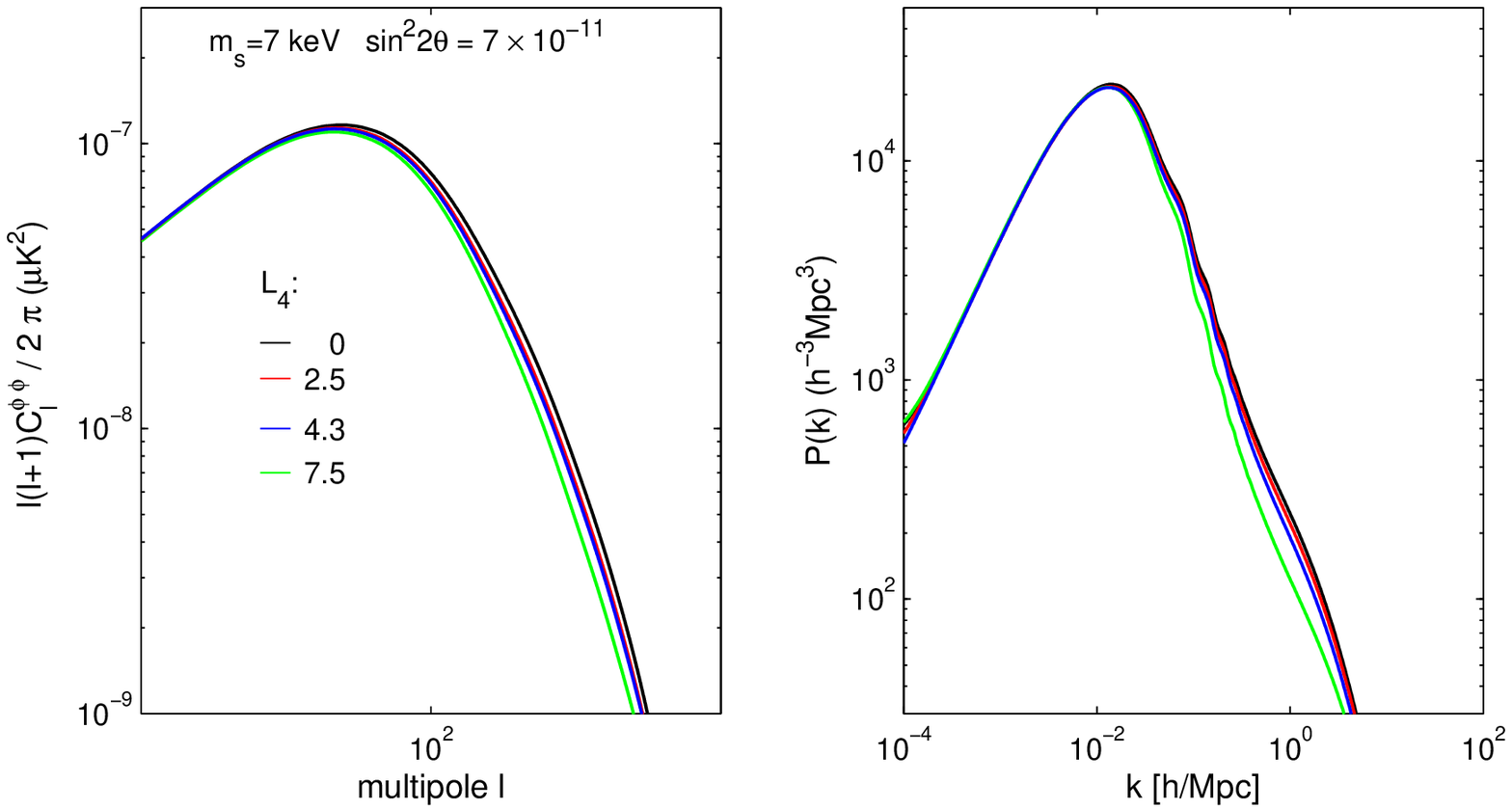}
\end{center}
\caption{The reconstructed gravitational lensing potential power spectrum $C^{\phi,\phi}_l$ (left) and the matter density fluctuations power spectrum $P(k)$ (right) for $m_s=7.14$ keV, 
$\sin^2 2 \theta$=6.46 $\times$10$^{-11}$  and
lepton asymmetry numbers $L_4$=2.5,4.3 and 7.5 having larger cutoff scales.} 
\end{figure}
We compute the angular power spectra of the CMB anisotropies, 
the matter density fluctuations power spectrum and the gravitational 
lensing potential power spectrum with  the Boltzmann Code for Anisotropies in the Microwave Background, CAMB  \cite{camb}, modified to allow 
the resonant flavor conversion production calculations as presented in the previous section.
The $\nu_{\alpha}$/${\bar \nu}_{\alpha}$  and $\nu_s$ phase-space distributions   
alter the homogeneous $\nu_{\alpha}$/${\bar \nu}_{\alpha}$ and $\nu_s$ energy density and pressure, the perturbed inhomogeneous components and the gravitational source term in the
Boltzmann equation that depends on the logarithmic derivative of the phase-space distributions with respect to comoving momentum.\\ 
The impact of the leptonic asymmetry on cosmological observables 
has been discussed in the literature \cite{Pastor1999,Ichiki,Popa2008}.
In particular, the non-zero leptonic asymmetry change
the neutrino velocity dispersion, the neutrino free-streaming length and the Jeans mass \cite{Hu95,Trotta2005}.
The leptonic asymmetry also increases 
the radiation energy density, usually parametrized by variation of the number of relativistic degrees of freedom $\Delta N_{eff}(\xi_{\nu})$ given by:
\begin{equation}
\label{delta_neff_zeta}
\Delta N_{eff} (\xi_{\nu})= 3 \left[\frac{30}{7} \left(\frac{\xi_{\nu}}{\pi}\right)^2
+\frac{15}{7} \left(\frac{\xi_{\nu}}{\pi}\right)^4 \right] \,.
\end{equation}
Figure~4 presents the gravitational lensing potential power spectrum $C^{\phi,\phi}_l$  and the matter density fluctuations power spectrum $P(k)$ for $m_s=7.14$ keV, 
$\sin^2 2 \theta$=6.46 $\times$10$^{-11}$ and 
lepton asymmetry numbers $L_4$=2.5,4.3 and 7.5, having larger cutoff scales.\\
The leptonic asymmetry shifts the beta equilibrium between protons and neutrons at the BBN epoch, leading to indirect effects on the CMB anisotropy through the primordial helium 
mass fraction $Y_P$ that decreases monotonically with increasing $\xi_{e}$.\\
The primordial helium mass fraction is also sensitive to the shape of 
$\nu_{e}/{\bar \nu}_{e}$ phase-space distribution.
As the  $\nu_e/{\bar{\nu}_e}$ phase-space distributions determine the rates of the 
neutron and proton interaction rates at BBN, the  
$\nu_{e}/{\bar \nu}_{e}$ spectra will change these rates and hence
the helium mass fraction over the case with thermal Fermi-Dirac spectrum \cite{Aba05,Kishimoto06,CSmith}.
We use the BBN PArthENoPE code \cite{Pisanti07, Kneller04} 
to compute the helium abundance, $Y_P$, for different values of $\Omega_b h^2$, 
$\Delta N_{eff}$, $\xi_{\nu}$ and changes of neutron and proton interaction rates.\\
We split the total variation of the radiation energy density in two uncorrelated  contributions:
\begin{equation}
\label{delta_neff}
\Delta N_{eff} = \Delta N_{eff}(\xi_{\nu})+ \Delta N^{oth}_{eff} \,,
\end{equation}
first due to the net leptonic asymmetry 
of the neutrino background  as given by Eq. (\ref{delta_neff_zeta})
and second due to possible extra contributions from unknown processes.

We adapt the latest version of the publicly available package CosmoMC \cite{Lewis02} 
for our cosmological analysis and use the following datasets and likelihood codes:
\begin{itemize}
\item The {\sc Planck} CMB anisotropy angular power spectrum, combined with WMAP-$9$ year polarization power spectrum at low $\ell$ \cite{WMAP9} and the corresponding codes \cite{Planck1,PlanckXV}:
\texttt{Commander}, that computes the low-$l$ {\sc Planck} likelihood,
\texttt{CamSpec}, that computes the {\sc Planck}  likelihood for  $50\leq l\leq 2500$,
\texttt{LowLike}, that computes the likelihoods for $2\leq l\leq 32$ polarization data
and \texttt{Lensing}, that computes the likelihoods from {\sc Planck} lensing power spectrum for $40\leq l\leq 400$ \cite{PlanckXVII}.
\item The high-$l$ CMB data from Atacama Cosmology Telescope(ACT) \cite{ACT13,Dunkley13} and the South Pole Telescope (SPT) \cite{Keisler11,Reichardt2013}. 
\item  The baryon acoustic oscillation (BAO) from the Sloan Digital Sky Survey (SDSS) Data Release 7 (DR7) \cite{BAODR7} at redshifts z= 0.2,0.35, the reanalyzed SDSS DR7 galaxy catalog data at z= 0.35 \cite{DR7, Padman12}, the SDSS Baryon Oscillation Spectroscopic Survey(BOSS) Data Release 9 (DR9) at z = 0.57 \cite{DR9,BAODR9} and the 6dF Galaxy Survey (6dFGS) \cite{6dF} BAO data at z=0.1 \cite{6dF, BAO6dF}.
\end{itemize}

\subsection{Analysis}

In this section we evaluate the impact of sterile neutrino
resonant production mechanism on the cosmological parameters. 
We consider the following extensions of the standard $\Lambda$CDM cosmological model:
\begin{itemize}
\item{\bf M1: $\Lambda$CDM+$\Sigma m_{\nu_{\alpha}}$}:\\
The minimal extension of the base $\Lambda$CDM model
by the addition of three degenerated species of massive neutrinos with total mass $\Sigma m_{\nu_{\alpha}}$
(within which $\Lambda$CDM is nested at $N_{eff}$=3.046 and $Y_P$=0.24).
\item{ \bf M2: $\Lambda$CDM+$\Sigma m_{\nu_{\alpha}}$ +$\Sigma \xi_{\nu_{\alpha}}$+$\Delta N_{eff}$ +$Y_P$}:\\
The addition to ${\bf M1}$ of the neutrino chemical potential $\Sigma \xi_{\nu_{\alpha}}$ 
the extra radiation energy density $\Delta N_{eff}$ as given in Eq. (\ref{delta_neff})
and the BBN prediction of the primordial helium abundance $Y_P$.
\item{ \bf M3: $\Lambda$CDM+$\Sigma m_{\nu_{\alpha}}$ +$\Sigma \xi_{\nu_{\alpha}}$+$\Delta N_{eff}$ +
$m_s$+$\sin^2 2\theta$+$Y_P$}:\\
The addition to ${\bf M2}$ of one DM sterile neutrino with the mass $m_s$, the matter mixing angle $\sin^2 2\theta$ and the BBN prediction of the primordial helium abundance $Y_P$.
\end{itemize}
Our main results are summarized in Table~\ref{tab:par}.
\begin{table}[tb]
\caption{The table shows the mean values and the absolute errors on the
main cosmological parameters obtained from the 	fits of different extensions
of the $\Lambda$CDM model discussed in the text with {\sc Planck}+WP+highL+BAO+lensing
dataset. For all parameters, except $\Sigma m_{\nu}$ and $\sin^2 2 \theta$
we quote the errors at $68\%$ CL. 
For  $\Sigma m_{\nu}$ and $\sin^2 2 \theta$
we give the values of $95\%$ upper limits.}

\vspace{0.3cm}
\centering
\begin{tabular}{lccc}
\hline\hline
Parameter         & M1 & M2 & M3 \\ \hline
$\Omega_bh^2$     & 0.02227$\pm$0.00019  &0.02221$\pm$0.00024  &0.02209$\pm$0.00023          \\ 
$\Omega_{cdm}h^2$ & 0.1161$\pm$0.0015    &0.1162$\pm$0.0015    &0.1169 $\pm$0.0021         \\
100$\theta_{MC}$  & 1.04156$\pm$0.00059  &1.04222$\pm$0.00095  &1.04261$\pm$0.00072       \\
$\tau$		  & 0.0882$\pm$0.0091    &0.0851$\pm$0.0092    &0.0814$\pm$0.0121	\\
$\Sigma m_{\nu_{\alpha}}$(eV)& $<$ 0.35  &$<$ 0.25&$<$ 0.21 \\
$m_s$(keV)& -& - & 6.08 $\pm$1.62\\
$\sin^2 2\theta$& - & - &$<$  5.61 $ \times 10^{-10}$ \\
$\xi_{\nu_{\alpha}}$& -                  & -0.062$\pm$0.079   &0.000002$\pm$0.000093 \\
$N_{eff}$& -                             & 3.37$\pm$0.19      &3.25$\pm$0.15\\
$n_s$&0.9586$\pm$0.0046                  &0.9679$\pm$0.0048   &0.9711$\pm$0.0052    \\
${\rm ln}10^{10}A_s$& 3.213$\pm$0.019    &3.231$\pm$0.023     &3.192$\pm$0.041 \\
\hline %----------------------------------------------------
$\Omega_{\Lambda}$& 0.699$\pm$0.012      &0.712$\pm$0.011     &0.713$\pm$0.012 \\
$\Omega_m$& 0.302$\pm$0.012              &0.288$\pm$0.011     &0.287$\pm$0.011 \\
$\sigma_8$&0.793$\pm$0.021&              0.801$\pm$0.022      &0.772$\pm$0.015  \\
$H_0$& 68.16$\pm$1.02                    &70.15$\pm$1.36      &69.95$\pm$1.17  \\
$Y_p$ & 0.24                             & 0.267$\pm$0.019    &0.24968$\pm$0.00011\\
$z_{eq}$& 3306$\pm$33                    & 3221 $\pm$36       &3183$\pm$46 \\
$L_{\nu_{\alpha}}$& -                    &-0.015$\pm$0.012    &0.000123$\pm$0.000022\\ 
$r_s/r_d$& 6.478$\pm$0.011               & 6.389$\pm$0.032    &7.013$\pm$0.011 \\ 
\hline\hline
\end{tabular}
\label{tab:par}
\end{table}

\subsubsection{Neutrino masses and mixing}

\begin{figure}
\label{prob1}
\begin{center}
\includegraphics[height=11cm,width=11cm]{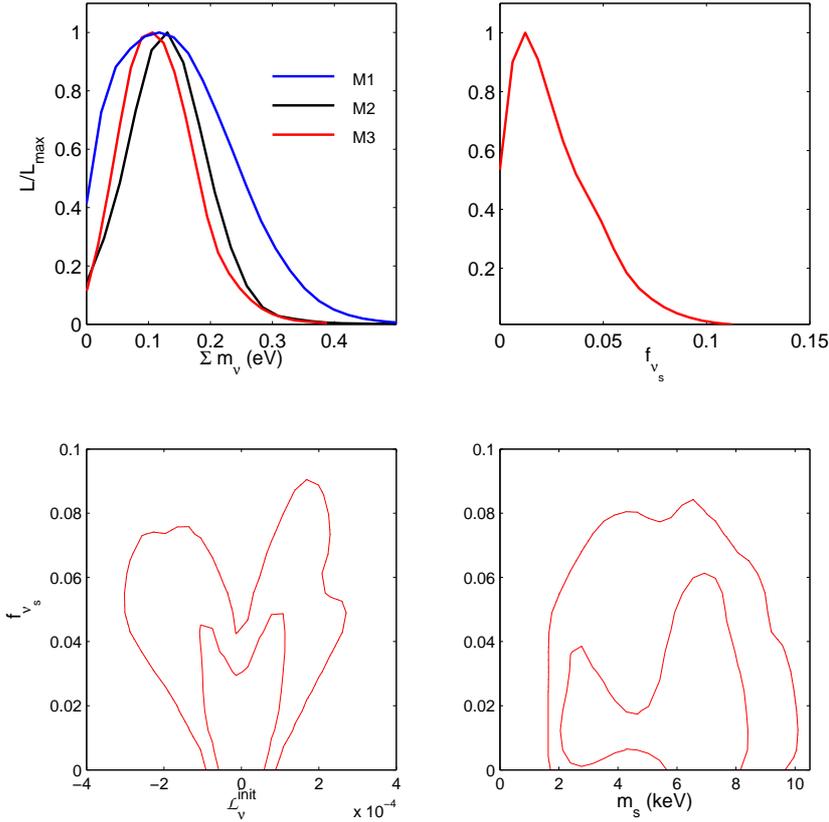}
\end{center}
\caption{Top: The marginalized likelihood posterior distributions for active neutrinos total mass $\Sigma m_{\nu_{\alpha}}$ and   
and sterile neutrino mass fraction, $f_{\nu_s}=\Omega_sh^2/\Omega_mh^2$,
as obtained from the fits of different extensions of the $\Lambda$CDM cosmological model with {\sc Planck}+WP+highL+BAO+lensing dataset
 (see also the text). 
Bottom: The joint confidence regions ${\cal L}^{init}_{\nu_{\alpha}}$ - $f_{\nu_{s}}$ 
and $m_s$ - $f_{\nu_{s}}$  (at 68\% CL and 95\% CL).}
\end{figure}
\begin{figure}
\label{prob2}
\begin{center}
\includegraphics[height=10cm,width=10cm]{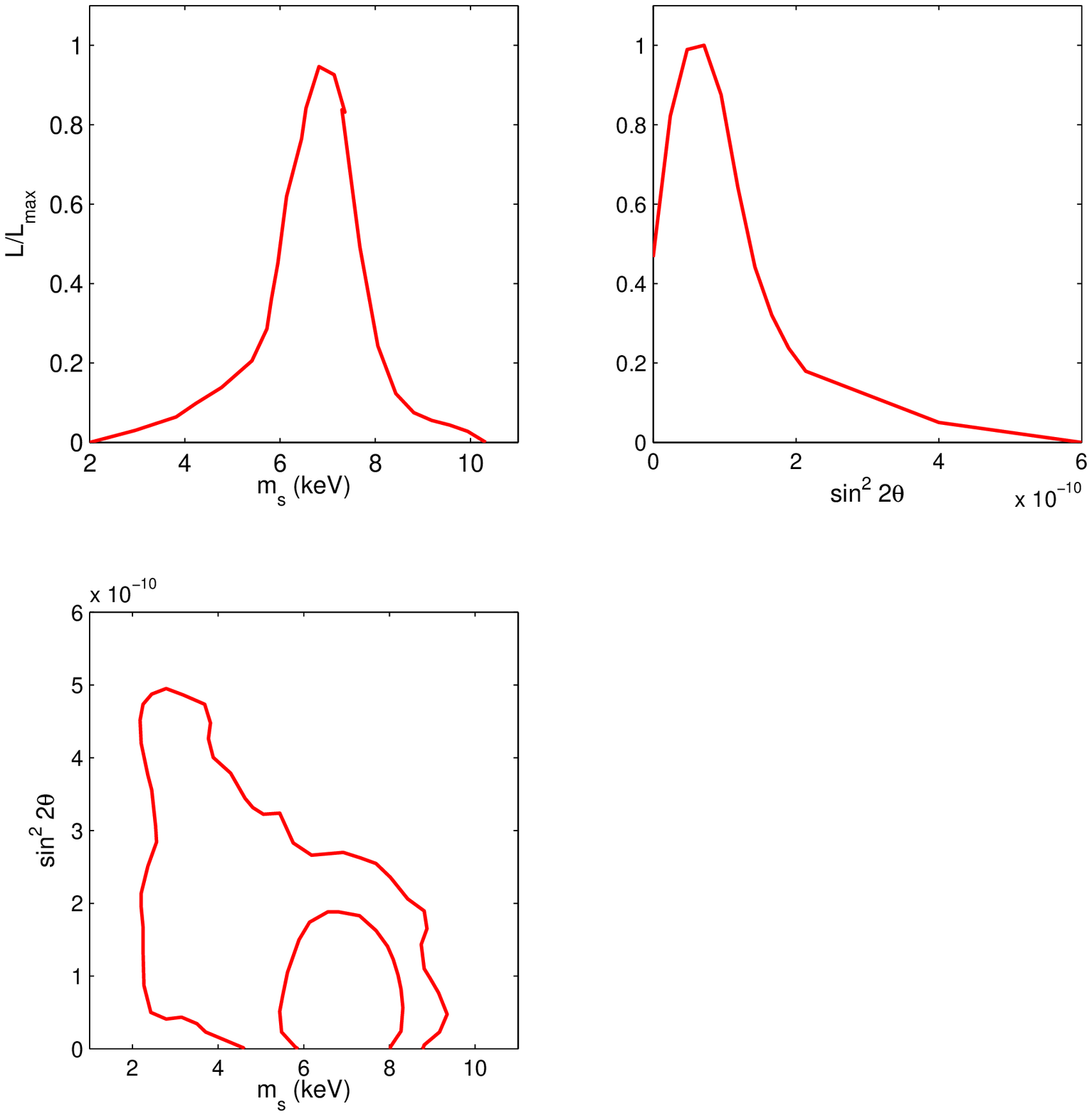}
\end{center}
\caption{Top: The marginalized likelihood posterior distributions obtained for 
the sterile neutrino mass, $m_s$ and the mixing angle $\sin^2 2\theta$.
Bottom: The joint confidence regions $m_s$ -$\sin^2 2 \theta$ (at 68\% CL and 95\% CL).}
\end{figure}
Top left panel from Figure~5 presents the likelihood posterior distributions of total active neutrino mass, $\Sigma m_{\nu_{\alpha}}$, 
showing a decrease of the upper limit (at 95\% CL), from 0.35 eV, for the case of minimal extension of the $\Lambda$CDM model (M1), to 0.25 eV when only the leptonic asymmetry is included (M2) and a subsequent decrease to 0.21 eV for the RP sterile neutrino scenario (M3).\\ 
In the top right panel of Figure~5 we plot the likelihood posterior probability distribution of the 
sterile neutrino mass fraction $f_{\nu_s}= \Omega_sh^2/\Omega_mh^2$.
The dominant effect on sterile neutrino resonant production  
is given by the value of the initial potential lepton number, ${\cal L}^{init}_{\nu}$, rather than sterile neutrino mass. This is shown in the bottom panels from Figure~5 that present 
the joint confidence regions ${\cal L}^{init}_{\nu_{\alpha}}$ - $f_{\nu_s}$ 
and $m_s$ - $f_{\nu_s}$.
We find $f_{\nu_s}<$ 0.078 for a value of 
lepton number per flavor $L_4 = 1.23 \pm 0.04$,
sterile neutrino mass $m_s =$6.08 $\pm$ 3.22 keV  
and mixing angle $\sin^2 2 \theta <$5.61 $\times$10$^{-10}$ (errors at 95\% CL).\\ 
Figure~6 presents the marginalized likelihood posterior distribution obtained for $m_s$ and the 
joint confidence regions $m_s$ -$\sin^2 2 \theta$.

\subsubsection{$^{4}$He abundance yield}

\begin{figure}
\label{epsL}
\begin{center}
\includegraphics[height=6cm,width=13cm]{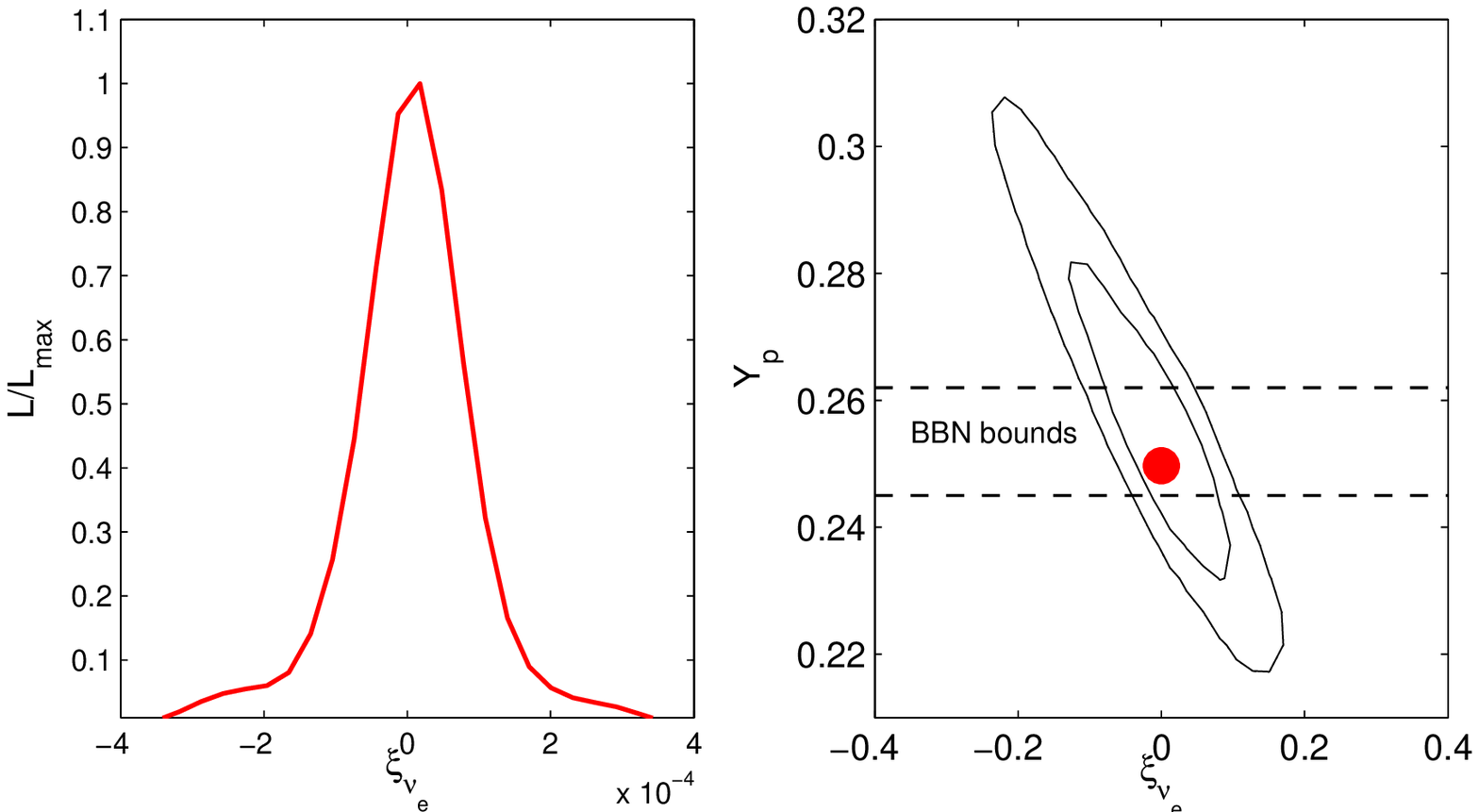}
\end{center}
\caption{Left: The marginalized likelihood probability distribution of
$\xi_{\nu_e}$ from the fit of model M3 
with  {\sc Planck}+WP+highL+BAO+lensing dataset.  
Right: The joint confidence regions $\xi_{\nu_e}$ - $Y_P$ (at 68\% CL and 95\% CL) 
from the fit of model M2 (black) and M3 (red) with the same dataset.
The latest observational bounds \cite{Aver} are also plotted (at 68\%CL).} 
\end{figure}
The BBN $^{4}$He yield depends sensitively  $\nu_{e}$/${\bar \nu}_{e}$ degeneracy parameter, $\xi_{\nu_e}$, as well as on the shape of its phase-space distribution function. 
We find the preference of cosmological data for smaller values of $\xi_{\nu_e}$ 
and $^{4}$He mass fraction, $Y_P$, in the case of RP sterile neutrino scenario (M3) when compared to the similar predictions obtained in the model with  leptonic asymmetry only (M2). 
This is shown in Figure~7 where 
we plot the marginalized likelihood probability distribution of
$\xi_{\nu_e}$ obtained from the fit of model M3 and the  joint confidence regions $\xi_{\nu_e}$ - $Y_P$ 
from the fit of model M2 and M3 with  our dataset.
We obtain $\xi_{\nu}$=0.0005$\pm$0.0004
and $Y_P$=0.2497 $\pm$ 0.0002 (95\% CL) for M3 case,  
in very good agreement with 
the {\sc Planck} determination \cite{Planck2}, $Y_P$ = 0.2485$\pm$0.0002, and 
the latest observational bounds $Y_P$=0.24968$\pm$0.00022 \cite{Aver}. 

\section{Conclusions}

\begin{figure}
\label{epsL}
\begin{center}
\includegraphics[height=10cm,width=10cm]{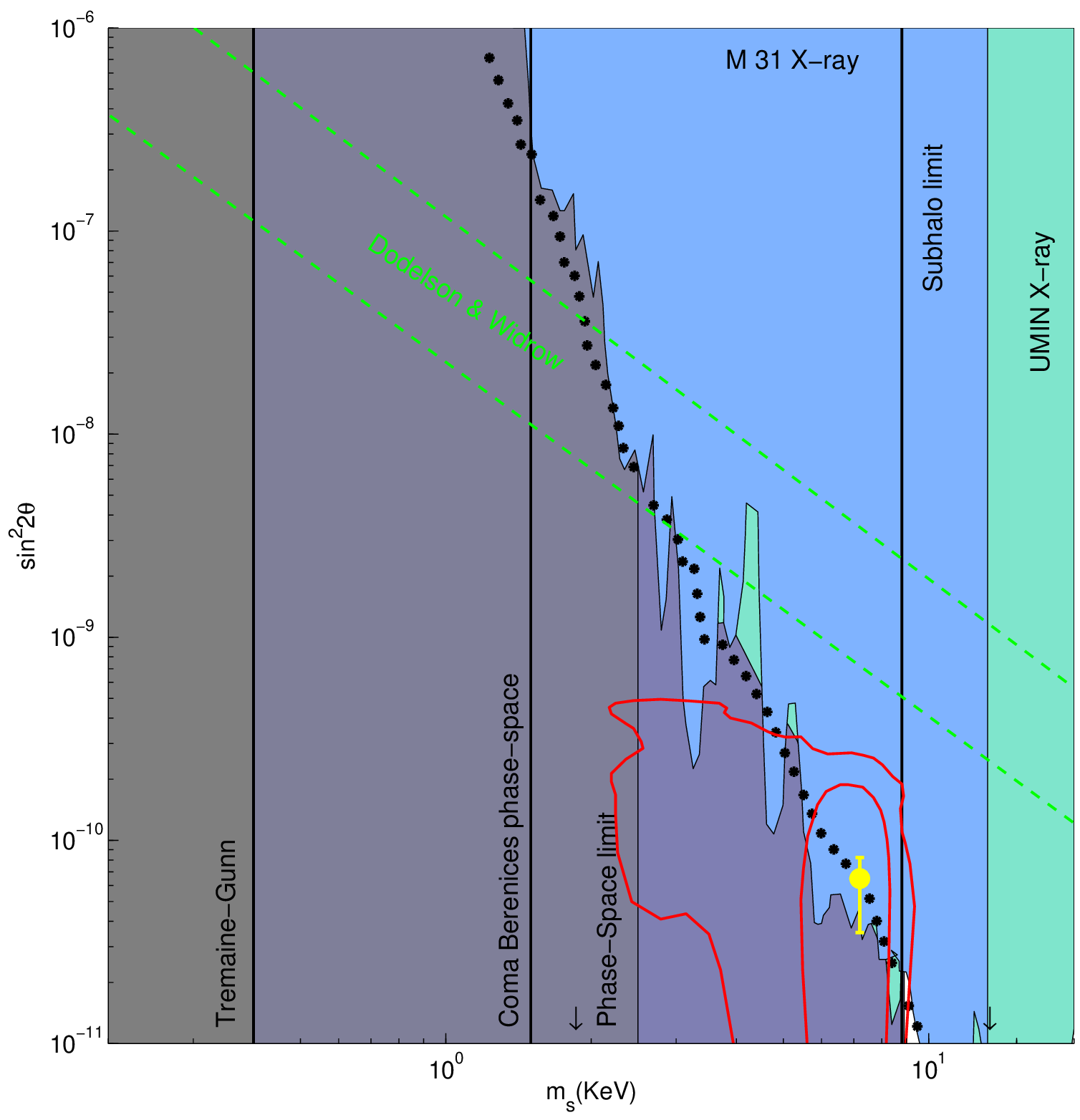}
\end{center}
\caption{Two-dimensional marginalized joint probability distributions at 68\% and 95\% CL (red lines) showing the dependence of RP 
sterile neutrino mass, $m_s$, and the matter mixing angle, $\sin^2 2 \theta$, obtained from our cosmological analysis.
We also show  the 95\% CL upper limits reproduced from Ref.\cite{Horiuchi14} based on deep Chandra \cite{Horiuchi14} 
and XMM-Newton \cite{Watson12} observations of M31 and Suzuku 
observations of Ursa Minor \cite{Loew,Urban}. 
The vertical lines 
represents the lower mass limits from Tremaine-Gunn phase-space considerations ($m_s \sim 0.4$) keV \cite{Gunn}, 
Coma Berenices phase-space ($m^{\rm DW}_s \sim 1.5$) keV, Segue I phase-space ($m^{\rm DW}_s \sim 2.5$) keV
and M31 subhalo counts ($m^{\rm DW}_s \sim 8.8$) keV. 
The upper and lower bounds (95\% CL) of  sterile neutrino mass obtained in Dodelson-Widrow model \cite{Asaka1} 
are also indicated (green dashed lines). 
The yellow symbol corresponds to the detection from Ref.\cite{Bulbul14a} ($m_s \simeq 7.1$  keV, $\sin^2 2\theta= 7 \times 10^{-11}$ 
and 90\% statistical error bar).} 
\end{figure}
%=====================================================================
Recently, few independent detections of a weak X-ray emission line at an energy of $\sim$ 3.5 keV seen toward a number of astrophysical sites have been reported.
Until now, none of these searches were able to establish precisely the origin of this line. 
If confirmed, this signal could be the signature of decaying DM sterile neutrino with a mass 
of $\sim 7.1$ keV.\\ 
The DM sterile neutrino production via resonant MSW conversion of 
active neutrinos to sterile neutrinos through the Shi-Fuller mechanism \cite{Shi99}
is the simplest model for the DM origin of the undefined $\sim$3.5 keV of X-ray line. \\
The RP sterile neutrino parameters required to produce this signal
have been recently inferred from the linear large scale structure
constraints to produce full DM density \cite{Abazajian14}.
These parameters are consistent with the Local Group and high-z galaxy count
constraints, fulfilling previously determined requirements to successfully
solve  ``missing satellite problem"and ``too-big-to-fail problem".
The cosmological constraints on subdominant RP sterile neutrino parameters
have not yet derived.

In this paper we place constraints on
RP sterile neutrino parameters in a $\Lambda$CWDM model containing a mixture of
WDM in the form of RP sterile neutrino and CDM, by using most of the present cosmological measurements.\\
We make a coupled treatment of the weak decoupling, primordial nucleosynthesis 
and photon decoupling epochs in the sterile neutrino RP scenario 
including the extra radiation energy density parametrized 
by the effective number of relativistic degrees of freedom $N_{eff}$.\\
We compute the radiation and matter perturbations including 
the full resonance sweep solution for $\nu_{\alpha}/{\bar \nu}_{\alpha} \rightarrow  \nu_{s}$ 
flavor conversion in the expanding Universe and provide constraints on the cosmological parameters 
and sterile neutrino properties, by using most of the present cosmological measurements.
In particular, the high precision CMB temperature anisotropy
and the reconstructed CMB gravitational lensing potential power spectra obtained 
by the {\sc Planck } satellite have impact on cosmological parameter degeneracies 
when DM sterile neutrino  scenario is considered.\\
We find that the values of the main cosmological parameters are in agreement with the 
predictions of the minimal extension of the base $\Lambda$CDM model except 
for the active neutrino total mass that is decreased to $\Sigma m_{\nu_{\alpha}}<$0.21 eV 
(at 95\% CL). \\
Figure~8 presents the $m_s$ - $\sin^2 2 \theta$ joint probability distributions obtained from our cosmological 
analysis, compared with the most of the existing similar constrains.
The cosmological measurements are in agreement  with the 
sterile neutrino RP scenario with the following parameters (errors at 95\%CL): $m_s =$6.28 $\pm$ 3.2 keV,  
mixing angle $\sin^2 2 \theta <$5.61 $\times$10$^{-10}$, lepton number per flavor
$L_4 = 1.23 \pm 0.04$ and sterile neutrino mass fraction $f_{\nu_s}< 0.078$.\\
Our results are in agreement 
with the sterile neutrino resonant production parameters
inferred in Ref. \cite{Abazajian14} from the linear large scale structure constraints
to produce full Dark Matter density.\\
This reflects the sensitivity of the 
high precision CMB observables to the physics that determines 
the neutrino energy spectra at different epochs, 
that require a self-consistent and coupled treatment \cite{Grohs15}. 
 
%===============================================================
\vspace{0.5cm}

{\bf Acknowledgments} \\\\
This work  was supported by a grant of the Ministery of National Education, CNCS-UEFISCDI,
project number PN-II-ID-PCE-2012-4-0511.

\end{document}